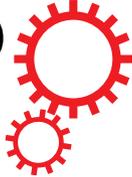
 

# Weak localization effect in topological insulator micro flakes grown on insulating ferrimagnet BaFe$_{12}$O$_{19}$

Guolin Zheng[1,*], Ning Wang[1,*], Jiyong Yang[1], Weike Wang[1], Haifeng Du[1], Wei Ning[1], Zhaorong Yang[1,2], Hai-Zhou Lu[3], Yuheng Zhang[1,2] & Mingliang Tian[1,2,4]

Many exotic physics anticipated in topological insulators require a gap to be opened for their topological surface states by breaking time reversal symmetry. The gap opening has been achieved by doping magnetic impurities, which however inevitably create extra carriers and disorder that undermine the electronic transport. In contrast, the proximity to a ferromagnetic/ferrimagnetic insulator may improve the device quality, thus promises a better way to open the gap while minimizing the side-effects. Here, we grow thin single-crystal Sb$_{1.9}$Bi$_{0.1}$Te$_3$ micro flakes on insulating ferrimagnet BaFe$_{12}$O$_{19}$ by using the van der Waals epitaxy technique. The micro flakes show a negative magnetoresistance in weak perpendicular fields below 50 K, which can be quenched by increasing temperature. The signature implies the weak localization effect as its origin, which is absent in intrinsic topological insulators, unless a surface state gap is opened. The surface state gap is estimated to be 10 meV by using the theory of the gap-induced weak localization effect. These results indicate that the magnetic proximity effect may open the gap for the topological surface attached to BaM insulating ferrimagnet. This heterostructure may pave the way for the realization of new physical effects as well as the potential applications of spintronics devices.

A gap opened for the surface states by breaking time reversal symmetry in topological insulators is anticipated to host many novel physics[1–8]. Experimentally, the gap may be realized either by magnetic doping[9–15], or by magnetic proximity to a ferromagnetic insulator[16–18]. One of the signatures of the gap openings is the weak localization effect[19]. The effect can give rise to positive low-field magnetoconductivity at low temperatures[19–22]. In contrast, for gapless surface states, a π Berry phase always leads to weak anti-localization and an associated negative magnetoconductivity[23–26]. However, in actual samples, the magnetic doping inevitably introduces magnetic scattering centers, defects, as well as magnetic clusters, which lead to mixed surface and bulk phases in magnetotransport[27,28]. As a result, it is hard to distinguish magnetically-doped topological insulators from diluted magnetic semiconductors[20], in the latter the weak localization-like magnetoconductivity is also anticipated and not attributed to the gap of the surface states. Compared to the magnetic doping, the magnetic proximity effect may also induce a gap for the surface states of topological insulator. A higher Curie temperature magnetic order can be achieved in a heterostructure of topological insulator and ferromagnetic insulator if the Curie temperature of the ferromagnetic insulator is high enough[29]. Moreover, the topological insulator-ferromagnetic insulator heterostructure is expected to suppress external magnetic impurities and magnetic clusters; therefore, it may be a better experimental candidate to induce the gap for the topological surface states. A number of heterostructures have been studied[29–34] with different ferromagntic insulator substrates, such as EuS[30,31], yttrium iron garnet[29,33], GdN[32] and BaFe$_{12}$O$_{19}$ (BaM)[34]. In the experiments, only a suppressed weak antilocalization effect with a negative magnetoconductivity was achieved. The suppressed weak antilocalization cannot be unambiguously attributed to gap opening because random magnetic scattering can also induce the suppression of the weak antilocalization effect[19,26]. A negative magnetoresistance in low fields has been demonstrated in a Bi$_2$Se$_3$-EuS heterostructure with a Bi$_2$Se$_3$ layer thinner than 4 nm[31]. However, it is not sufficient to conclude that the magnetic proximity has indeed opened the gap since the finite-size effect can also open gaps in thin films[35] and leads to the weak localization effect[22]. Very recently, a low-field positive magnetoconductivity was observed in a Bi$_2$Se$_3$-BaM heterostructure in parallel magnetic fields[34], but the perpendicular magnetoconductivity remains negative. Domain walls may be the possible origins of the positive parallel-field magnetoconductivity in very weak parallel fields. Most heterostructures have been fabricated by the molecular beam epitaxy (MBE) method, and the size of these heterostructures is much larger than the magnetic domains of the ferrimagnetic insulator. Thus massive Dirac electrons would be expected in magnetic domain areas, but remain massless at the domain walls[29]. The domain walls may suppress conductivity[28]; then a positive magnetoconductivity arises as the magnetic field removes the domain walls.

[1]High Magnetic Field Laboratory, the Chinese Academy of Sciences, Hefei 230031, the People's Republic of China; University of Science and Technology of China, Hefei 230026, The People's Republic of China. [2]Collaborative Innovation Center of Advanced Microstructures, Nanjing University, Nanjing 210093, The People's Republic of China. [3]Department of Physics, South University of Science and Technology of China, Shenzhen, China. [4]Hefei Science Center, Chinese Academy of Sciences, Hefei 230031, Anhui, China. *These authors contributed equally to this work. Correspondence and requests for materials should be addressed to H.-Z.L. (email: luhz@sustc.edu.cn) or M.T. (email: tianml@hmfl.ac.cn)

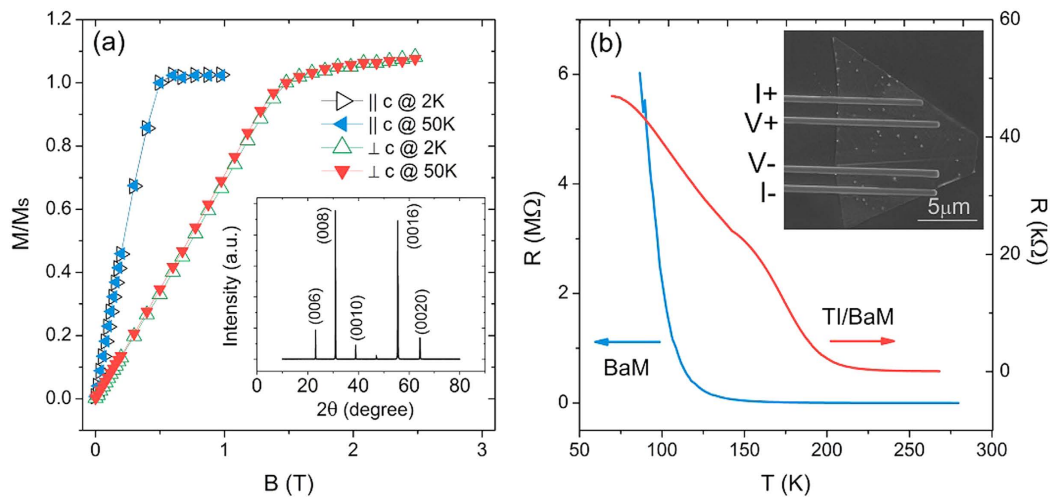

**Figure 1. Device characteristics.** (**a**) The magnetic moments of the single crystal ferromagnetic insulator BaFe$_{12}$O$_{19}$ (BaM). The out-of-plane and in-plane magnetic moments are indicated by "$\parallel$ c" and "$\perp$ c", respectively. The magnetic moments in the two directions do not change as the temperature increases from 2 K to 50 K. Inset: the XRD pattern of the single crystal BaM. Only (00$l$) peaks related to the hexagonal phase can be observed. (**b**) The R-T curves of the BaM substrate only and the heterostructure of topological insulator and BaM, respectively. Inset: The scanning electron microscope image of the Sb$_{1.9}$Bi$_{0.1}$Te$_3$-BaFe$_{12}$O$_{19}$ heterostructure, with the current (I+ and I-) and voltage (V+ and V-) probes. The white points are redundant tellurium particles generated during cooling. The warping edges show the large lattice mismatch between Sb$_{1.9}$Bi$_{0.1}$Te$_3$ and BaFe$_{12}$O$_{19}$. The scale bar is 10 $\mu$m.

In this work, we fabricated BaFe$_{12}$O$_{19}$-Sb$_{1.9}$Bi$_{0.1}$Te$_3$ heterostructures by using the van der Waals epitaxial technique. The size of the topological insulator flakes can be controlled to be comparable with the magnetic domains of BaM. In perpendicular magnetic fields, a positive magnetoconductivity possibly associated with the weak localization effect is observed, indicating that the magnetic proximity has opened a gap for the surface state of topological insulator. The parallel-field magnetoconductivity shows a negative magnetoconductivity near zero field as the in-plane magnetization of BaM is not able to open the gap for the surface states. Using the magnetoconductivity formula for the competition between weak antilocalization and weak localization effects, we fitted the magnetoconductivity curves in perpendicular fields and found that the surface gap $\Delta$ induced by the magnetic proximity is about 10 meV. Our results demonstrate that the magnetic proximity can break time-reversal symmetry and open a sizable gap for the surface states of topological insulator. This topological insulator-BaM heterostructure thus may pave the way for further experimental research on novel physics and potential applications of spintronics devices.

## Results

**Heterostructures of topological insulator and BaFe$_{12}$O$_{19}$.** Hexagonal BaFe$_{12}$O$_{19}$ is a well-known ferrimagnetic insulator with a uniaxial anisotropy along the c crystallographic axis. The magnetic domain structure of a single-crystal BaM has been verified with positive and reversed magnetic domains along the c axis under different directions of magnetization. The size of the magnetic domains is about 5 $\mu$m. The domains exhibit labyrinth, stripe, honeycomb-type patterns as the magnetization field tilts from the c axis to the a-b plane[36,37]. We prepared the BaM single crystals by the floating zone method. We chose large and flat single crystals with a natural cleavage plane (0001) as the ferrimagnetic insulator substrate. Figure 1(a) shows the magnetization of the BaM substrate measured by the Magnetic Property Measurement System, where M is the magnetic moment, and M$_S$ is the saturation magnetic moment. The saturation magnetization field H$_S$ of BaM along the in-plane (perpendicular to the c axis) and out-of-plane directions are 1.5 T and 0.5 T, respectively. There is no obvious change of H$_S$ in both directions when the temperature increases from 2 K up to 50 K. Due to the free moving of domain walls in response to the variations of the external fields, no evident hysteresis is observed at low fields[38].

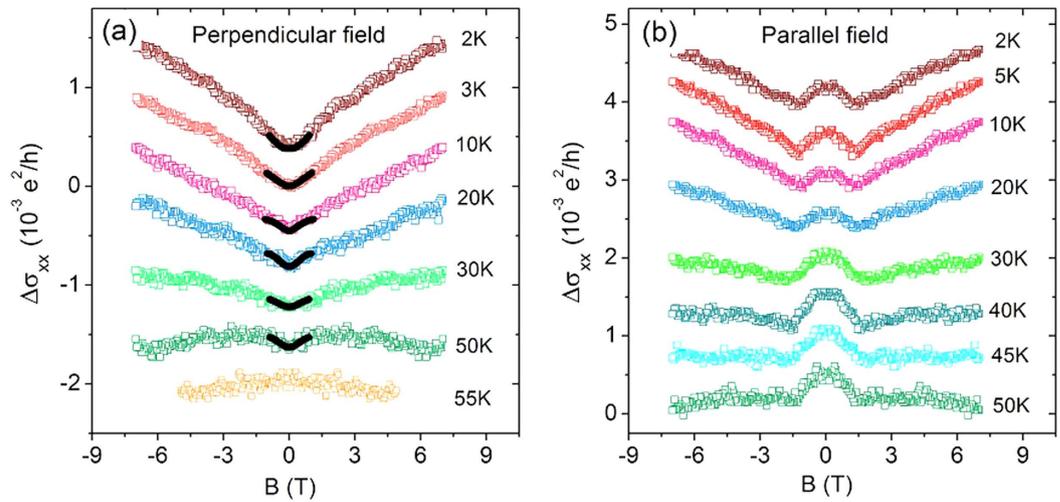

**Figure 2. Magnetoconductivity of heterostructures.** (**a**) In perpendicular magnetic fields. (**b**) In parallel magnetic fields. The solid curves within 1 T in (**a**) are the fitting curves by using Eq. (2). The data curves at different temperatures are offset for clarity.

The van der Waals epitaxy is a facile method to grow high-quality nanostructures on a clean surface of substrate irrespective of large lattice mismatch[39,40]. By this method, we successfully fabricated $Sb_{1.9}Bi_{0.1}Te_3$-BaM heterostructures in a 1-inch horizontal tube furnace via the catalyst-free vapor-solid (v-s) growth technique similar to that in ref. 41. We choose the stoichiometric $Sb_{1.9}Bi_{0.1}Te_3$, because it can effectively lift the position of the Dirac point out of the bulk valence band while tuning the Fermi level inside the bulk gap through charge compensation[42]. The inset of Fig. 1(b) presents the scanning electron microscope (SEM) image of the $Sb_{1.9}Bi_{0.1}Te_3$ nanoplate on the BaM substrate. The warping edges of the nanoplate indicate the large lattice mismatch between $Sb_{1.9}Bi_{0.1}Te_3$ and BaM. The white points on the $Sb_{1.9}Bi_{0.1}Te_3$ nanoplate are redundant tellurium generated during the cooling. Figure 1(b) shows the R-T curves of both the heterostructure and BaM substrate. BaM is a ferrimagnetic insulator with high room temperature resistance. After the growth of the topological insulator on BaM, the resistance of BaM was reduced to 200 Ω at 300 K, revealing that the BaM becomes conductive after annealing for 1 hour near 300 °C. When the temperature decreased to 125 K, the resistance of BaM increased sharply. In contrast, the resistance of the heterostructure increased slowly with a decrease of temperature. For instance, at T = 100 K, the resistance of the BaM substrate reached above $2 \times 10^6$ Ω, which is 100 times larger than the resistance of the heterostructure. This high resistance indicates that when T < 100 K, the current mainly flows through the topological insulator film. Here, our temperature range of interest is below 55 K, at which the BaM substrate becomes a full insulator.

**Magnetoconductivity and weak localization.** We measured the magnetoconductivity of our samples by the standard four-probe transport measurement. Figure 2(a) shows the magnetoconductivity of the $Sb_{1.9}Bi_{0.1}Te_3$-BaM heterostructure at different temperatures in perpendicular magnetic fields. We observed a positive magnetoconductivity (i.e., negative magnetoresistance) in the perpendicular field up to 7 T at low temperatures. The positive magnetoconductivity weakened as the temperature increased and finally the magnetoconductivity displayed mixed behavior from positive in low fields to negative in high fields when the temperature increased up to 50 K. In the perpendicular field, the classical magnetoconductivity arising from the Lorentz force always gives a negative magnetoconductivity (positive magnetoresistance). In the strong disorder regime, the transport can exhibit a negative magnetoresistance[43], but it requires that the conductivity $\sigma \ll e^2/h$. While at 2 K, the conductivity $\sigma$ in our heterostructures ranges between $e^2/h < \sigma < 10\ e^2/h$, not in the strong disorder regime. The positive magnetoconductivity here is most likely the consequence of the weak localization effect of the gapped surface states and, therefore, has a quantum interference origin[19]. As the temperature increased to T = 55 K, the magnetoconductivity became entirely negative. There is no obvious change in the magnetization of our BaM substrates near 50 K, so the observed sign change of the magnetoconductivity near 50 K is not due to the gap closing with vanishing magnetization, but possibly because the temperature quenched the quantum interference and then the classical magnetoconductivity became overwhelming. In order to conduct a comparative study, we also grew $Sb_{1.9}Bi_{0.1}Te_3$ on a $SiO_2$ substrate. As shown in Fig. 3(a), the perpendicular-field magnetoconductivity of $Sb_{1.9}Bi_{0.1}Te_3$ on the $SiO_2$ substrate exhibited a sharp cusp near zero field. The cusp is a typical signature of the weak antilocalization effect for the topological surface states due to their nature as massless Dirac fermions.

**Magnetization induced surface state gap.** Functionally, the magnetoconductivity of the weak antilocalization and weak localization effects can always be fitted by the Hikami-Larkin-Nagaoka (HLN) formula[44],

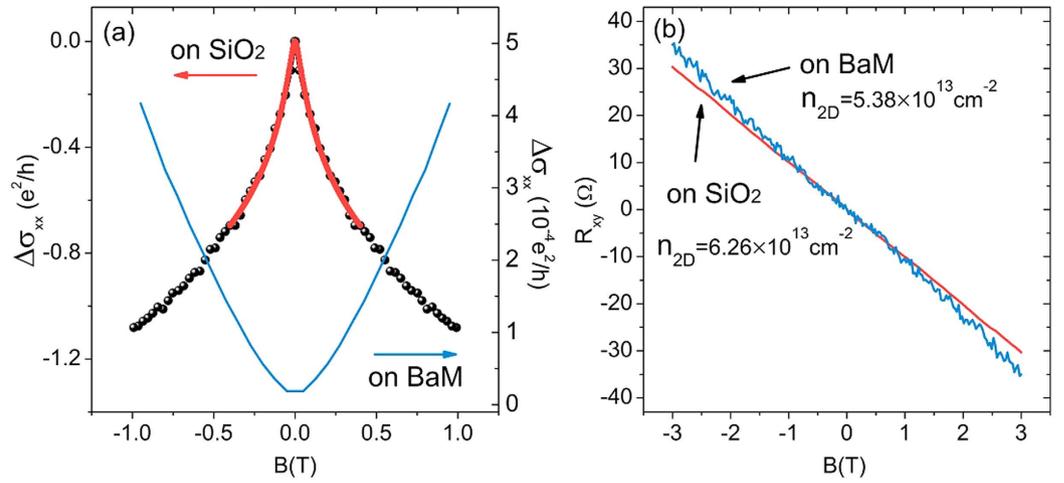

**Figure 3. Transport properties of the control sample.** (**a**) The comparison of the magnetoconductivity in perpendicular fields between the control sample (topological insulator on SiO$_2$) and the topological insulator on BaM at 2 K. The black curve is the experimental data while the red curve is the fitting. The fitting yields $l_\phi = 135$ nm. (**b**) The Hall resistance of the topological insulator grown on the SiO$_2$ and BaM substrates, respectively.

$$\Delta\sigma(B) = -\alpha \frac{e^2}{\pi h}\left[\psi\left(\frac{1}{2} + \frac{\hbar}{4el_\phi^2}\right) - \ln\left(\frac{\hbar}{4el_\phi^2}\right)\right], \quad (1)$$

where $\psi$ is the digamma function. The prefactor $\alpha$ and the phase coherence length $l_\phi$ serve as two fitting parameters. However, the information about the gap opening cannot be given by the HLN formula. To quantitatively study the gap opening in the surface states, we employed the magnetoconductivity formula derived for the weak antilocalization-localization crossover of the massive Dirac fermions[19],

$$\Delta\sigma(B) = \sum_{i=0,1} \frac{\alpha_i e^2}{\pi h}\left[\psi\left(\frac{1}{2} + \frac{l_B^2}{l_{\phi i}^2}\right) - \ln\left(\frac{l_B^2}{l_{\phi i}^2}\right)\right], \quad (2)$$

where the magnetic length $l_B^2 = \hbar/4e|B|$, $1/l_{\phi i}^2 = 1/l_\phi^2 + 1/l_i^2$, and $\alpha_0$, $\alpha_1$, $l_0^2$, $l_1^2$ are explicit functions (see Methods) of $\Delta/2E_F$, with $\Delta$ the gap of surface states and $E_F$ the Fermi level measured from the Dirac point. In the weak antilocalization limit, $\alpha_0 = 0$, $\alpha_1 = -0.5$, while in the weak localization limit, $\alpha_0 = 0.5$, $\alpha_1 = 0$. Equation (2) is valid when the mean free path is much shorter than the magnetic length ($l_e^2 \ll l_B^2$). We used formula (2) to fit the magnetoconductivity in perpendicular fields. The solid lines in Fig. 2(a) are the fitting curves by using equation (2). Figure 4(a) shows the fitting result of $\Delta/2E_F$ as a function of temperature. At T = 2 K, $\Delta/2E_F$ is about 0.329. With increasing temperature, $\Delta/2E_F$ decreases. At T = 50 K, $\Delta/2E_F = 0.304$, changed by less than 10%. These values of $\Delta/2E_F$ are in the regime between the unitary and orthogonal symmetry classes, where a suppressed weak localization effect can give rise to a positive magnetoconductivity[19]. In the earlier experiments[45,46], the higher Fermi energies in topological insulators Bi$_2$Se$_3$ and Bi$_2$Te$_3$ may greatly reduce $\Delta/2E_F$, leading to only weak antilocalization-like magnetoconductivity according to the theory[19].

In order to evaluate the surface gap, we used 2D carrier density $n_{2D}$ derived from the Hall measurement to estimate the Fermi level. By measuring the Hall resistance of two similar devices on the SiO$_2$ and BaM substrates, respectively, as shown in Fig.3 (b), we obtain that $n_{2D} = 6.26 \times 10^{13} e^{-1} cm^{-2}$ on SiO$_2$ and $n_{2D} = 5.38 \times 10^{13} e^{-1} cm^{-2}$ on the BaM substrate. The values of the carrier density means that the Fermi energy is at the bottom of the bulk conduction band, which is about 0.1–0.2 eV from the Dirac point of the surface states, according to the ARPES data in ref.42. We estimate that the surface state gap induced by the magnetic proximity is of the order of

$$\Delta = 2E_F/3 \approx 10 \text{ meV}. \quad (3)$$

As shown in Fig. 4(b), with increasing temperature, the fitted phase coherence length is reduced by half, from 28 nm at 2 K to 14 nm at 50 K, showing a much stronger temperature dependence compared to that of $\Delta/2E_F$. Therefore, the fitting results indicate that the reduction of the phase coherence length is the main reason why the positive magnetoconductivity of weak localization is suppressed in Fig. 2.

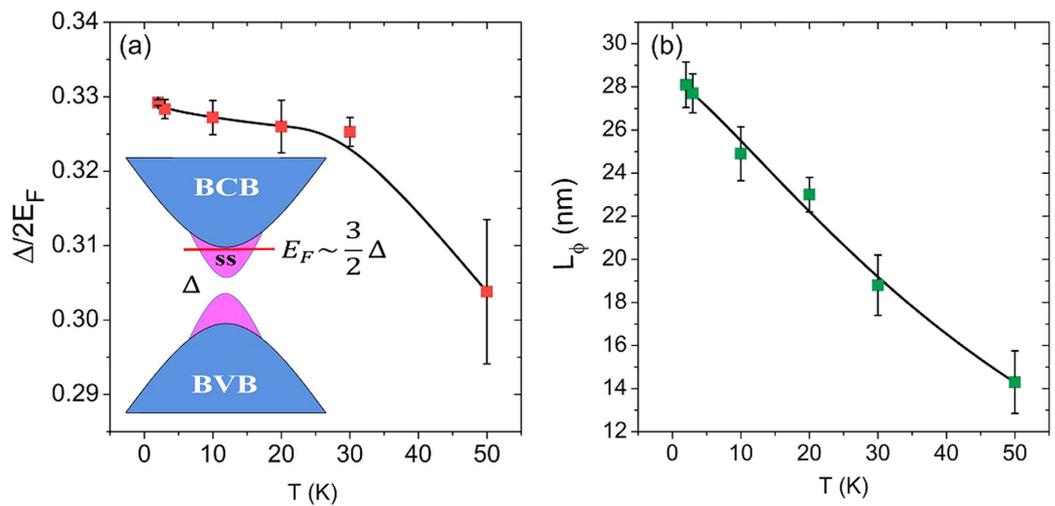

**Figure 4. Fitting results of the magnetoconductivity in perpendicular magnetic fields.** (**a**) The fitted $\Delta/2E_F$ as a function of temperature, where $\Delta$ is the gap of the surface states and $E_F$ is the Fermi level. Inset: a schematic illustration of the band structure of the topological insulator on BaM. The bulk conduction band, bulk valence band, and surface states are indicated by BCB, BVB, and SS, respectively. (**b**) The fitted phase coherence length, which is suppressed with increasing temperature. The relative change of $\Delta/2E_F$ with temperature is much smaller than that of the phase coherence length. The fitting is performed by using Eq. (2).

## Discussion

We also show the magnetoconductivity data in parallel fields in Fig. 2(b). The positive magnetoconductivity can still be recognized in the high-field range, but a deflection appears near 1.5 T and the magnetoconductivity becomes negative near the zero field, as shown in Fig. 2(b). The negative magnetoconductivity near the zero field indicates that there is no gap opening for the surface states in parallel magnetic fields. This absence of gap is because while applying the external magnetic field in the in-plane direction, the spin of BaM substrate will be rotated to the in-plane direction, and correspondingly, the surface state of topological insulator remains gapless. As the temperature goes up to 30 K, the deflection of the magnetoconductivity curves is weakened and disappears at 50 K. This deflection behavior has also been observed in magnetically doped topological insulators[47] and undoped ultrathin topological insulator films[22,48,49]. In theory, if the magnetoconductivity curves exhibit a deflection, the magnetoconductivity are contributed by multiple channels and the bulk states might contribute to the transport besides the surface states[50,51]. In topological insulator ultrathin films, this can be well explained by the 2D modified Dirac model[35], while in our heterostructure, the thickness of the topological insulator nanoflakes is about 20 nm, which is much thicker than the ultrathin limit (<5 nm). Besides, in the parallel fields, the observed deflection in magnetoconductivity happens at 1.5 T, at which the weak-field limit (usually within 0.5 T) for the weak antilocalization and weak localization effects has been violated. The observed magnetoconductivity in parallel fields cannot be explained by the competition between weak antilocalization and weak localization. The mechanism for the deflection in the parallel-field magnetoconductivity is not clear so far. One possible explanation is similar to that in ref.28, where domain walls suppress the conductivity while the magnetic field may remove the domain walls gradually, leading to the positive magnetoconductivity at high fields. The size of our heterostructure is comparable to the single magnetic domain in BaM, but the observed deflection cannot be explained by this mechanism. Another possible explanation is related to the canting of magnetization as discussed in refs.29,30. When the parallel field is larger than 1.5 T, the magnetization of BaM saturates in the in-plane direction. A canting of magnetization in the out-of-plane direction might form at the interface of our heterostructure due to strong spin-orbit coupling as well as a large anisotropy of the Lande g factor for the interface electrons[29,30]. The surface state gap might also be opened by the canting of magnetization in the high-field range. Also, we can see in Fig. 2, the $\Delta\sigma$ induced by the magnetization canting is much smaller than that in the perpendicular field direction.

The magnetic proximity effect usually is short-range, especially for a material with weak spin-orbit interaction. If the ferrimagnetic substrate cannot affect the top surface, there should be no apparent change in the resistance. However, in our Fig. 1(b), the resistance increases by one order, as the $SiO_2$ substrate is replaced with the ferrimagnet $BaFe_{12}O_{19}$. Considering the resistance of $BaFe_{12}O_{19}$ is several orders higher, the resistance increase is not due to $BaFe_{12}O_{19}$ alone, but most likely from the topological insulator thin film under the influence of a long-range magnetic proximity effect. One of the possible reasons for the long-range proximity effect is strong spin-orbit coupling. It has been shown that the magnetic proximity effect can induce an obvious negative magnetoresistance in a Platinum film attached to a ferrimagnetic insulator YIG, up to a film thickness of 14 nm[52]. The topological insulator $Sb_{1.9}Bi_{0.1}Te_3$ also has strong spin-orbit coupling, so the magnetic proximity between $Sb_{1.9}Bi_{0.1}Te_3$ and $BaFe_{12}O_{19}$ might be also long-range. In the device for the Hall-bar meassurement in Fig. 3(b), the thickness of the topological insulator is 30 nm. In the Hall-bar device, we did not observe the negative magnetoresistance, probably because of poor interface contact or thicker topological insulator layer.

## Methods

**Synthesis of heterostructure.** The mixed powder of $Sb_2Te_3$ and $Bi_2Te_3$ with molar ratio 19:1 was placed in the hot center of the tube furnace. Hexagonal BaM single crystals with a clean cleavage plane (0001) were put at a location about 9–12 cm away from the hot center. The tube is cleaned by ultrapure Ar gas for five times prior to the growth, and then the Ar gas flow is fixed at 50 s.c.c.m. In order to avoid the escape of tellurium, the hot centre was heated to 475°C within 5 min and kept at this temperature for one hour, followed by natural cooling. During the whole process, we maintained the pressure in the tube at 2.8 Torr. We obtained high-quality van der Waals epitaxy heterostructures of topological insulator and ferromagnetic insulator only when the substrate was placed 11.5 cm away from the hot center.

**Device fabrication and characterization.** To make standard four-probe devices on the topological insulator-BaM heterostructure for transport measurement, we transferred the heterostructure into a FEI SEM/FIB dual beam system for the deposition of Pt electrodes. The four Pt-stripe electrodes have a nominal width of $0.8\,\mu m$ and a thickness of 200 nm deposited from a precursor gas of (methylcyclopentadienyl) trimethyl platinum $(CH_3)_3CH_3C_5H_4Pt$. To avoid Pt contamination and Ga ion irradiation, we designed the distance between two adjacent electrodes to be larger than $2\,\mu m$. During the whole fabrication process, we only used the electron beam to guide the Pt deposition.

**Magnetoconductivity fitting formula.** We use equation (2) to fit the magnetoconductivity in perpendicular magnetic fields. The parameters in equation (2) are defined as[19]

$$\begin{aligned}
l_0^2 &= l_e^2 a^4 (a^4 + b^4 - a^2 b^2)[b^4(a^2 - b^2)^2]^{-1}, \\
l_1^2 &= l_e^2 (a^4 + b^4)^2 [a^2 b^2 (a^2 - b^2)^2]^{-1}, \\
\alpha_1 &= -a^4 b^4 [(a^4 + b^4)(a^4 + b^4 - a^2 b^2)]^{-1}, \\
\alpha_0 &= (a^4 + b^4)(a^2 - b^2)^2 [2(a^4 + b^4 - a^2 b^2)^2]^{-1},
\end{aligned}$$

where $l_e$ is the mean free path, $a = \cos(\theta/2)$, $b = \sin(\theta/2)$, and $\cos\theta = \Delta/2E_F$. We assume that the effects of random magnetic scattering is to reduce the phase coherence length and have been included in the definition of $l_\phi$.

### Acknowledgements
We thank Murong Lang for fruitful discussion. This work was supported by the Natural Science Foundation of China (Grant No. 11174294, 11374302, 11574127, U1432251 and U1332209), and the program of Users with Excellence, the Hefei Science Center of CAS and the CAS/SAFEA international partnership program for creative research teams of China.

### Author Contributions
G.Z., N.W., H.Z.L. and M.T. conceived and designed the experiment. G.Z. and N.W., synthesized the samples and fabricated the devices. W.N., H.D., J.Y., W.W. and Z.Y. carried out the transport measurement. H.Z.L., G.Z., M.T. and Y. Z. provided the analysis and the theoretical interpretation. G.Z., H.Z.L. and M.T., wrote the paper, with assistance from all the co-authors.

### Additional Information
**Competing financial interests:** The authors declare no competing financial interests.

**How to cite this article**: Zheng, G. *et al.* Weak localization effect in topological insulator micro flakes grown on insulating ferrimagnet $BaFe_{12}O_{19}$. *Sci. Rep.* **6,** 21334; doi: 10.1038/srep21334 (2016).